\documentclass{article}
\usepackage[utf8]{inputenc}

\usepackage{amsmath,amsfonts,amssymb}
\usepackage{graphicx}
\usepackage{caption}
\usepackage{array}
\usepackage{subcaption}
\usepackage{bbm}
\usepackage{authblk}
\usepackage{float}
\usepackage{makecell}
\setcellgapes{4pt}
\title{Estimation of conditional mixture Weibull distribution with right-censored data using neural network for time-to-event analysis
}
\author{Achraf Bennis  \and
Sandrine Mouysset  \and
Mathieu Serrurier  }

\begin{document}
\maketitle
\thispagestyle{empty}
\pagestyle{empty}
\begin{abstract}
In this paper, we consider survival analysis with right-censored data which is a common situation in predictive maintenance and health field. We propose a model based on the estimation of two-parameter Weibull distribution conditionally to the features. To achieve this result, we describe a neural network  architecture and the associated loss functions that takes into account the right-censored data. We extend the approach to a finite mixture of two-parameter Weibull distributions. We first validate that our model is able to precisely estimate the right parameters of the conditional Weibull distribution on synthetic datasets.
In numerical experiments on two real-word datasets (METABRIC and SEER), our model outperforms the state-of-the-art methods. We also demonstrate that our approach can consider any survival time horizon.  
\end{abstract}

\section{INTRODUCTION}
\hspace{5mm}Time-to-event analysis, also called survival analysis, is needed in many areas. This branch of statistics which emerged in the $20^{th}$ century is heavily used in engineering, economics and finance, insurance, marketing, health field and many more application areas. Most previous works and diverse literature approach time-to-event analysis by dealing with time until occurrence of an event of interest; e.g. cardiovascular death after some treatment intervention, tumor recurrence, failure of an aircraft air system, etc. The time of the event may nevertheless not be observed within the relevant time period, and could potentially occur after this recorded time, producing so called right-censored data. The main objective of survival analysis is to identify the relationship between the distribution of the time-to-event distribution and the covariates of the observations, such as the features of a given patient, the characteristics of an electronic device or a mechanical system with some informations concerning the environment in which it must operate. The Weibull distribution could be used as lifetime distributions in survival analysis where the goal would be to estimate its parameters taking account the right-censored data. Several previous works focused on the estimation of a Weibull distribution with right-censored data (see  Bacha and Celeux \cite{bacha1996bayesian}, Ferreira and Silva \cite{ferreira2017parameter}, Shuo-Jye Wu \cite{wu2002estimations}, etc.)

Among the first estimators widely used in this field is the Kaplan-Meier estimator \cite{kaplan1958nonparametric} that may be useful to estimate the probability that an event of interest occurs at a given point in time. However, it is limited in its ability to estimate this probability adjusted for covariates; i.e. it doesn't incorporate observations' covariates. The semi-parametric Cox proportional hazards (CPH) \cite{COX72} is used to estimate covariate-adjusted survival, but it assumes that the subject's risk is a linear function of their covariates which may be too simplistic for many real world data. 
Since neural networks can learn nonlinear functions, many researchers tried to model the relationship between the covariates and the times that passes before some event occurs, including Faraggi-Simon network \cite{faraggi1995neural} who proposed a simple feed-forward as the basis for a non‐linear proportional hazards model to model this relationship. After that, several works focused on combining neural networks and survival analysis, notably DeepSurv \cite{katzman2016deep} whose architecture is deeper than Faraggi-Simon's one and minimizes the negative log Cox partial likelihood with a risk not necessarily linear. These models use multi-layer perceptron that is capable to learn non-linear models, but it is sensitive to feature scaling which is necessary in data preprocessing step and has limitations when we use unstructured data (e.g. images). There is a number of other models that approach survival analysis with right-censored data using machine learning, namely RandomForest Survival \cite{ishwaran2008random}, dependent logistic regressors \cite{NIPS2011_4210} and Liao's model \cite {liao2016combining} who are capable of incorporating the individual observation's covariates.

This paper proposes a novel approach to survival analysis: we assume that the survival times distribution are modeled according to a finite Mixture of Weibull distributions (at least one), whose parameters depends on the covariates of a given observations with right-censored data. As Luck \cite{luck2017deep}, we propose a deep learning model that learns the survival function, but we will do this by estimating the Weibull's parameters. Unlike DeepHit \cite{lee2018deephit} whose model consists on discretizing the time considering a predefined maximum time horizon. Here, as we try to estimate the parameters, we can model a continuous survival function, and thus, estimate the risk at any given survival time horizon. For this purpose, we construct a deep neural network model considering that the survival times follow a finite mixture of two-parameter Weibull distributions. This model, which we call \textbf{DeepWeiSurv} tries to estimate the parameters that maximizes the likelihood of the distribution. To prove the usefulness of our method, we compare its predictive performance with that of state-of-the art methods using two real-world datasets. DeepWeiSurv outperforms the previous state-of-the-art methods. 

\section{Weibull Mixture Distribution for survival analysis} 
\subsection{Survival Analysis with right-censored data}
\begin{figure}
    \centering
    \includegraphics[scale=0.6]{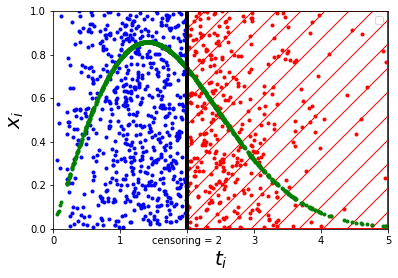}
    \centering
    \caption{Weibull distribution right-censored at $t_c=2$  with $x\in [0,1]$ uniformly distributed. In this figure, the parameters of the law are independent with regard to $x$.}
    \label{illus_data}
\end{figure}
Let $X = \{(x_i,t_i,\delta_i)| i\leq n\}$ be a set of observations with $x_i \in \mathbb{R}^d$, the $i^{th}$ observation of the baseline data (covariates), $t_i \in \mathbb{R}$ its survival time associated, and $\delta_i$ indicates if the $i^{th}$ observation is censored ($\delta_i = 0$) or not ($\delta_i = 1$). As can be seen in Figure \ref{illus_data}, a blue point represents an uncensored observation $(x_i, t_i,\delta_i=1)$ and a red point represents a censored observation $(x_i, t_i,\delta_i=0)$.  
In order to characterize the distribution of the survival times $T = (t_i|x_i)_{i\leq n}$, the aim is to estimate, for each observation, the probability that the event occurs after or at a certain survival time horizon $t_{STH}$ defined by: $$S(t_i|x_i) = P(t_i \geq t_{STH}|x_i).$$ Note that, $t_{STH}$ may be different to the censoring threshold time $t_{c}$. An alternative characterization of the distribution of $T$ is given by the hazard function $\lambda(t)$ that is defined as the event rate at time $t$ conditional on survival at time $t$ or beyond. Literature has shown that $\lambda(t)$ can be expressed as follows: $\lambda(t) = \frac{f(t)}{S(t)}$, $f(t)$ being the density function. \\

 Instead of estimating the $S(t_i|x_i)$, it is common to estimate directly the survival time $\hat{t}_i$. In this case, we can measure the quality of estimations with the concordance index \cite{harrell1982evaluating} defined as follows: 
\begin{equation}
    C_{index} = \frac{\sum_{i,j} \mathbbm{1}_{t_i > t_j}. \mathbbm{1}_{\hat{t}_i > \hat{t}_j}.\delta_j} {\sum_{i,j} \mathbbm{1}_{t_i > t_j}.\delta_j}. \label{Cindex}
\end{equation}
$C_{index}$ is designed to calculate the number of concordant pairs of observations among all the comparable pairs $(i,j)$ such that $\delta_i$=$\delta_j$=1. It estimates the probability $\text{P}$($\hat{t}_i > \hat{t}_j | t_i > t_j$) that compares the rankings of two independent pairs of survival times $t_i$,$t_j$ and associated predictions $\hat{t}_i$,$\hat{t}_j$.
\subsection{Weibull distribution for censored data}
\hspace{5mm}From now, we consider that $T$ follows a finite mixture of two-parameter Weibull (at least a single Weibull) distributions independently from $x_i$ (i.e. $S(t_i | x_i) = S(t_i)$). In this case, we have the analytical expressions of $S$ and $\lambda$  with respect to the mixture parameters. This leads to consider a problem of parameters estimation of mixture of Weibull distributions with right-censored observations. 

\subsubsection{Single Weibull case}
Here, we are dealing with a particular case where $T$ follows a single two-parameter Weibull distribution, $W$($\beta$, $\eta$), whose parameters are $\beta > 0$ (\textit{shape}) and $\eta > 0$ (\textit{scale}). We can estimate these parameters by solving the following likelihood optimization problem:
\begin{equation*}
 (\hat{\beta},\hat{\eta}) = \underset{\beta,\eta}{\mathrm{argmax}} 
 \{\mathcal{LL}(\beta, \eta | (t_i,\delta_i)_i\} = \sum_{i=1}^{n} \delta_{i}log[(S_{\beta,\eta}.\lambda_{\beta,\eta})(t_i)] + (1-\delta_i)log[S_{\beta,\eta}(t_c)]
\end{equation*} 
where: \begin{equation*}
\begin{split}
    S_{\beta,\eta}(y) &= exp[-(\frac{y}{\eta})^\beta],\\
    \lambda_{\beta,\eta}(y) &= (\frac{\beta}{\eta})(\frac{y}{\eta})^{\beta-1}
\end{split}
\end{equation*} 
and $t_c$ being the censoring threshold time. $\mathcal{LL}$ is the log-likelihood of Weibull distribution with right-censored data. To be sure that the $\mathcal{LL}$ is concave, we make a choice to consider that the shape parameter $\beta$ is greater than 1 ($\beta \ge 1$). 
\subsubsection{Mixture case.}
Now, we suppose that $T$ follows $\mathcal{W}_p$ = {\Large [}{\large (}${W}(\beta_k, \eta_k)${\large )}, {\large (}$\alpha_k${\large )}{\Large ]}$_{k=1..p}$ a mixture of $p$ Weibull distributions with its weighting coefficients ($\sum_{k} \alpha_k$ = 1, $\alpha_k$ $\ge$ 0). In statistics, the density associated is defined by: 
\begin{equation*}
    f_{\mathcal{W}_p}=\sum_k \alpha_k f_{\beta_k,\eta_k}\underset{}{\mathrm{=}}\sum_k \alpha_k S_{\beta_k,\eta_k}\lambda_{\beta_k,\eta_k}.
\end{equation*} 
Thus, the log-likelihood of $\mathcal{W}_p$ can be written as follows:
\begin{equation}\begin{split}
    \mathcal{LL}(\beta,\eta,\alpha | y) = \sum_{i=1}^{n} \delta_i log\begin{bmatrix}\sum_k \alpha_k (S_{\beta_k,\eta_k}.\lambda_{\beta_k,\eta_k})(t_i)\end{bmatrix}  \\
    + (1-\delta_i)log\begin{bmatrix}\sum_k \alpha_k S_{\beta_k,\eta_k}(t_{c})\end{bmatrix}.
    \end{split}
    \label{loss_mix}
    \end{equation}
In addition to the mixture's parameters $(\beta_k, \eta_k)_{k=1..p}$, we need to estimate the weighting coefficients $(\alpha_k)$ considered as probabilities. Therefore, we estimate the tuple $(\alpha, \beta, \eta)$ by solving the following problem:
\begin{equation*}
    (\hat{\beta},\hat{\eta},\hat{\alpha}) = \underset{\beta,\eta,\alpha}{\mathrm{argmax}} \{\mathcal{LL}(\beta,\eta,\alpha |(t_i, \delta_i)_i\}
\end{equation*}
Knowing Weibull's mean formula $\mu$ and given that the mean of a a mixture is a weighted combination of the means of the distributions that form this mixture (more precisely, $\mu$ = $\sum_k \alpha_k \mu_k$), the mean lifetime can thus be estimated as follows:
\begin{equation}
    \mu= \hat{\alpha}.diag(\hat{\eta}).\Gamma(1 + \frac{1}{\hat{\beta}})^T
    \label{mu}
\end{equation}
where $\Gamma$ is the Gamma function. $\mu$ can be used as the survival time estimation for the computation of the concordance index (with $\hat{t_i}=\mu_i=\mu$ when the parameters of the distribution are independent from $x_i$). 
\section{Neural network for estimating conditional Weibull mixture}
We now consider that the Weibull mixture's parameters depend on the covariates $x$=$(x_i)$. We propose to use a neural network to model this dependence.
\subsection*{Model description}
\begin{figure}
    \centering
    \includegraphics[scale=0.6]{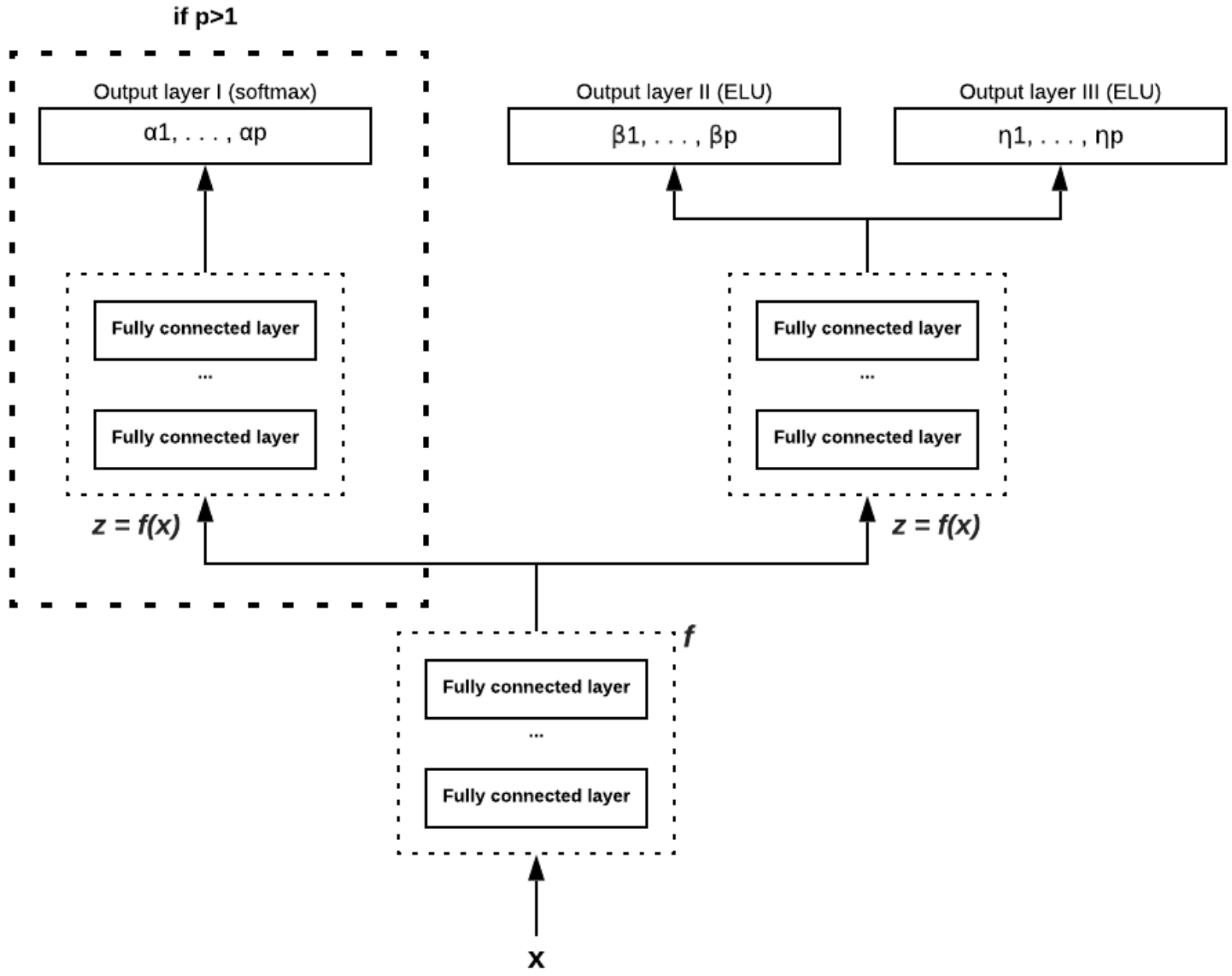}
    \centering
    \caption{The architecture of DeepWeiSurv}
    \label{DeepWeiSurvArchi}
\end{figure}

We name $g_p$ the function that models the relationship between $x_i$ and the parameters of the conditional Weibull mixture:

\begin{equation*}
\begin{array}{ccccc}
g_p & : & \mathbb{R}^d & \to &  \mathbb{R}^{p\times 3} \\ 
& & x_i & \mapsto &  (\alpha, \beta, \eta) 
\end{array}
\end{equation*}
where $\alpha = (\alpha_1,..\alpha_p)$ and $(\beta,\eta) = {\Large (}(\beta_1,..,\beta_p),(\eta_1,..,\eta_p){\Large )}$.
Note that, when $p=1$, it is no more required to estimate $\alpha$.
This function is represented by the network named DeepWeiSurv described in Figure \ref{DeepWeiSurvArchi}. Hence, our goal is to train the network to learn $g_p$ and thus $(\hat{\beta},\hat{\eta})$ the vector of parameters that maximise the likelihood of the time-to-event distribution ($\hat{\alpha}$ as well if $p>1$). DeepWeiSurv is therefore a multi-task network. It consists of a common sub-network, a classification sub-network (\textit{clf}) and a regression sub-network (\textit{reg})
The shared sub-network takes as an input the baseline data \textbf{x} of size $n$ and compute a latent representation of the data $z$. When $p>1$, \textit{clf} and \textit{reg} take $z$ as an input towards producing $\hat{\alpha} $ and  $(\hat{\beta},\hat{\eta})$ respectively. For \textit{reg} sub-network, we use ELU (with its constant = 1) as an activation function for both output layers. We use this function to be sure that we have enough gradient to learn the parameters thanks of the fact that it becomes smooth slowly unlike ReLU function. However the codomain of ELU is $]-1,\infty[$, which is problematic given the constraints on the parameters mentioned in the previous section ($\beta \geq 1$ and $\eta > 0$). To get around this problem, the network will learn $\hat{\beta}_{off}=\hat{\beta}+2$ and $\hat{\eta}_{off}=\hat{\eta}+1+\epsilon$. The offset is then applied in the opposite direction to recover the parameters concerned. For the classification part we need to learn $\alpha \in \mathbb{R}^p$. To ensure that $\sum_k \alpha_k=1$ and $\alpha_k\in [0,1]$, we use a $softmax$ activation in the output layer of \textit{clf}. 
For each $ 1 \le k \le p$, \textit{clf} produces,  $\alpha_k = (\alpha_{1k},..\alpha_{nk})$ where $\alpha_{ik}$ is such that: $\hat{P}(\{Y = t_i\}) = \alpha_{ik}$ with $Y \sim W(\beta_{k}, \eta_k)$ and $\hat{P}$ a probability estimate, whereas $reg$ outputs $\beta_k$ = $(\beta_{1k},..\beta_{nk})$ and $\eta_k$ = $(\eta_{1k},..\eta_{nk})$. Otherwise, i.e. $p$ = 1, we have $\alpha_1$ = 1, thus we don't need to train \textit{clf}. \\
\begin{figure}[ht]
    \centering
    
    \includegraphics[scale=0.37, width=7cm]{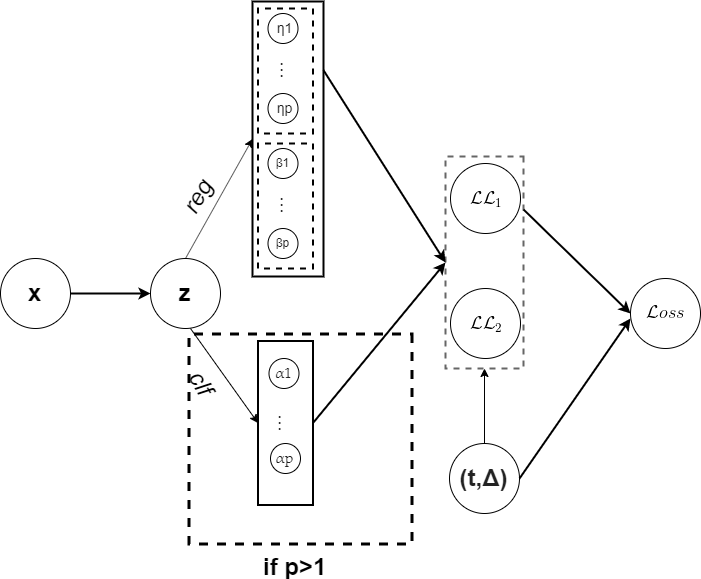}
    \centering
    \caption{Computational graph of $\mathcal{L}oss$}
    \label{lossgraph}
\end{figure}
To train DeepWeiSurv, we minimize the following loss function: $$\mathcal{L}oss = - \mathcal{LL}(\beta,\eta,\alpha | y) \underset{(\ref{loss_mix})}{=} \mathcal{LL}_1 . \Delta^T + \mathcal{LL}_2.(\textbf{1}_{\mathbb{R}^n}-\Delta)^T,$$ where $\Delta$ is the vector of event indicators and:
\begin{equation*}
    \mathcal{LL}_1 = log{\large [}\hat{\alpha}.S\Lambda_{\hat{\beta},\hat{\eta}}(T){\large ]}
 \text{ and }
\mathcal{LL}_2 = log{\large [}\hat{\alpha}.S_{\hat{\beta},\hat{\eta}}(t_c){\large ]}
\end{equation*}
with:
\begin{equation*}
S\Lambda_{\hat{\beta},\hat{\eta}}(t) = \begin{pmatrix}
(S_{\beta_1,\eta_1}.\lambda_{\beta_1,\eta_1})(t_1) & ... & (S_{\beta_1,\eta_1}.\lambda_{\beta_1,\eta_1})(t_n) \\
... & ... & ... \\ 
(S_{\beta_p,\eta_p}.\lambda_{\beta_p,\eta_p})(t_1) & ... & (S_{\beta_p,\eta_p}.\lambda_{\beta_p,\eta_p})(t_n)
\end{pmatrix}\end{equation*} \text{and }
\begin{equation*}
S_{\hat{\beta},\hat{\eta}}(t_c) = \begin{pmatrix}
S_{\beta_1,\eta_1}(t_c) \\ ...  \\ S_{\beta_p,\eta_p}(t_c)\end{pmatrix}
\end{equation*}

$\mathcal{LL}_1$ exploits uncensored data, whereas $\mathcal{LL}_2$ exploits censored observations by extracting the knowledge that the event will occur after the given censoring threshold time $t_c$. Figure \ref{lossgraph} is an illustration of the computational graph of our training loss: the inputs are the covariates x, the real values of time and event indicator $(t,\Delta)$ and the outputs are the estimates $(\hat{\alpha},\hat{\beta},\hat{\eta})$. 
\subsection*{Experiment on SYNTHETIC dataset}
\begin{figure}
    \centering
    \includegraphics[width=0.45\textwidth]{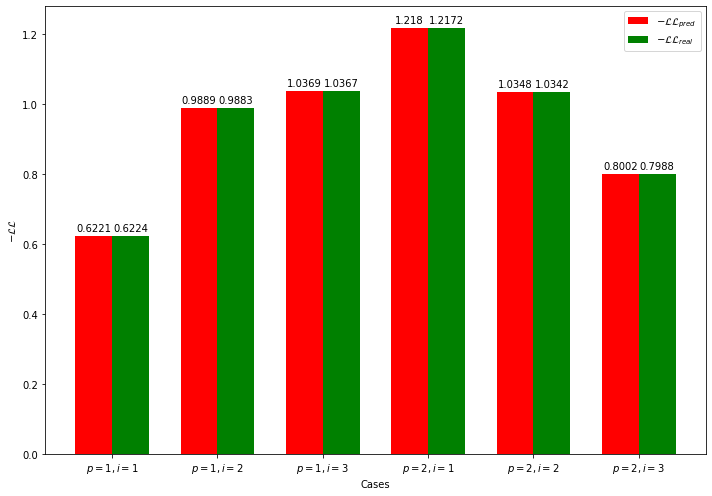}
    \centering
    \caption{$-\mathcal{LL}_{pred}$ and $-\mathcal{LL}_{real}$ values for each studied case}
    \label{bar_synth}
\end{figure}
The main objective in this section is to validate mathematically DeepWeiSurv, that is, to show that this latter is able to estimate the parameters. For this purpose, we perform an experiment on a simulated data. In this experiment, we treat the case of a \textit{single} Weibull distribution ($\alpha_{p=1}$ = $1$) and a \textit{mixture} of 2 Weibull distributions ($\alpha_{p=2}$ = $(0.7,0.3)$) using three different functions: $f_1$ (\textit{linear}), $f_2$ (\textit{quadratic}), $f_3$ (\textit{cubic}). For each function $f_i$ we generate $T_{p=1}^i$ $\sim$ $W(\beta_1^i, \eta_1^i)$ and $T_{p=2}^i$ $\sim \mathcal{W}_{p=2}$($\beta_{0.7}^i, \eta_{0.7}^i, \beta_{0.3}^i, \eta_{0.3}^i$). We compare the predicted likelihood with the real, and optimal one. These two likelihoods are equal when the estimated parameters correspond to the real ones.
Let $X$ be a vector of 10000 observations generated from an uniform distribution $\mathcal{U}_{[0,1]}$. Here we select 50\% of observations to be right censored at the median of survival times $t_m$ ($\delta_i$ = 0 if $t_i > t_m$). 
We set the parameters to be the following functions:

\begin{equation*}
\begin{pmatrix}
\beta_1^1 \\ 
\eta_1^1\\
\beta_{0.7}^1\\
\eta_{0.7}^1\\
\beta_{0.3}^1\\
\eta_{0.3}^1 
\end{pmatrix} = \begin{pmatrix}
 3 & 2  \\ 
 2 & 1\\ 
 2 & 1  \\ 
 1 & 2 \\
 1 & 2 \\
 3 & 1 
\end{pmatrix}.\begin{pmatrix}
X \\
1
\end{pmatrix}
\end{equation*}

\begin{equation*}
\begin{pmatrix}
    \beta_1^2 \\
    \eta_1^2 \\
    \beta_{0.7}^2\\
    \eta_{0.7}^2\\
    \beta_{0.3}^2\\
    \eta_{0.3}^2
    \end{pmatrix}
    = \begin{pmatrix}
     2 & 1 & 1  \\ 
     1 & 2 & 1 \\
     2 & 2 & 1 \\
     1 & 3 & 1 \\
     1 & 1 & 2 \\
     1 & 0 & 2
    \end{pmatrix}.\begin{pmatrix}
    X^2\\
    X\\
    1
    \end{pmatrix}
\end{equation*}

\begin{equation*}
    \begin{pmatrix}
\beta_1^3 \\
\eta_1^3 \\
\beta_{0.7}^3\\
\eta_{0.7}^3\\
\beta_{0.3}^3\\
\eta_{0.3}^3
\end{pmatrix}=\begin{pmatrix}
2 & 0 & 1 & 1 \\
1 & 1 & 0 & 1 \\
2 & 0 & 1 & 1 \\
1 & 1 & 0 & 1 \\
1 & 2 & 0 & 1 \\
3 & 2 & 0 & 1
\end{pmatrix}.\begin{pmatrix}
X^3\\
X^2\\
X\\
1
\end{pmatrix}
\end{equation*}

The bar plot in Figure \ref{bar_synth} displays the predicted likelihood $-\mathcal{LL}_{pred}$ of each distribution and their real one $-\mathcal{LL}_{real}$. We notice that the real value and predicted one of each case are very close to each together which means that the model can identify very precisely the parameters of the conditional distributions. Now, we test DeepWeisurv on the real-world datasets. 
\section{Experiments}
We perform two sets of experiments based on real survival data : METABRIC and SEER. We give a brief descriptions of the datasets below; Table \ref{Descriptive} gives an overview on some descriptive statistics of both real-word datasets. We train DeepWeiSurv on real survival datasets. We compare the predictive performance of DeepWeiSurv with that of CPH\cite{COX72} which is the most-widely used model in the medical field and DeepHit\cite{lee2018deephit} that seems to achieve outperformance over previous methods. These models are also tested in the same experimental protocol as DeepWeiSurv.
\subsubsection{\textbf{METABRIC}}
METABRIC (Molecular Taxonomy of Breast Cancer International Consortium) dataset is for a Canada-UK project that aims to classify breast tumours into further subcategories. It contains gene expressions profiles and clinical features used for this purpose. In this data, we have 1981 patients, of which 44.8\% were died during the study and 55.2\% were right-censored. We used 21 clinical variables including tumor size, age at diagnosis, Progesterone Receptor (PR) status etc (see Bilal et al. \cite{bilal2013improving}).
\subsubsection{SEER}
The Surveillance, Epidemiology, and End Results (SEER\footnote{https://seer.cancer.gov})\cite{SEERNov18} Program provides information on cancer statistics during 1975-2016. We focused on the patients (in total 33387) recorded between 1998 and 2002 who died from a breast cancer BC (42.8\%) or a heart disease HD (49.6\%), or who were right-censored (57.2\% and 50.4\% respectively). We extracted 30 covariates including gender, race, tumor size, number of malignant of benign tumors, Estrogen Receptor status (ER), PR status, etc. For evaluation we separated the data into two datasets with respect of the death's cause (BC \& HD) while keeping censored patients in both of them.
\begin{table}
\centering
\captionsetup{justification=centering}
\begin{tabular}{|c|c|c|c|c|}
\hline
{\textbf{Datasets}}& {No. Uncensored} & {No. censored} & \multicolumn{2}{c|}{No. Features}\\
& & & Qualitative & Quantitative \\
\hline
 METABRIC& 888 (44.8\%)&1093 (55.2\%) &15 &6 \\
\hline
SEER BC&9152(42.8\%) & 12221 (57.2\%)& 23& 11\\

SEER HD&12014 (49.6\%) &12221 (50.4\%) &23 &11 \\
   \hline 
\end{tabular}
\caption{Descriptive Statistics of Real-World Datasets}
\label{Descriptive}
\end{table}
\subsubsection*{Network Configuration}
DeepWeiSurv is consisted of three blocks: the shared sub-network which is a 4-layer network, 3 of which are fully connected layers (128, 64, 32 nodes respectively) and the remain is a batch normalization layer, the second and the third block (\textit{reg}, \textit{clf} respectively) consisted of 2 fully connected layers (16, 8 nodes) and 1 batch normalization layer. Added to that, the network finishes by one softmax layer and two ELU layers as outputs. The hidden layers are activated by ReLU function. DeepWeiSurv is trained via Adam optimizer and learning rate of $10^{-4}$. DeepWeiSurv is implemented in a PyTorch environment. 
\subsubsection{Experimental Protocol}
We applied 5-fold cross validation: the data is randomly splitted into training set (80\% and 20\% of which is reserved for validation) and test set (20\%). We use the predicted values of the parameters to calculate the mean lifetime $\mu$ and then $C_{index}$  defined by equation (\ref{Cindex}). This latter is calculated on the validation set. We tested DeepWeiSurv with $p = 1$ and $p=2$ (we tested higher values of $p$, but without better performances). 
\subsubsection{Results} 
\begin{table}
\centering
\captionsetup{justification=centering}
\begin{tabular}{ | c | c | c | c |}
\hline

Algorithms& {METABRIC} & {SEER BC} & {SEER HD}\\ \hline \hline

 CPH &0.658&0.833&0.784\\
   & (0.646 - 0.671) & (0.829 - 0.838) & (0.779 - 0.788)\\
\hline
DeepHit&0.651&0.875&0.846\\
   & (0.641 - 0.661) & (0.867 - 0.883) & (0.842 - 0.851)\\\hline 
DeepWeiSurv($p$ = $1$)&0.805&0.877&0.857\\
   & (0.782 - 0.829) & (0.864 - 0.891) & (0.85 - 0.866)\\\hline
DeepWeiSurv($p$ = $2$)&\textbf{0.819}&\textbf{0.908}&\textbf{0.863}\\
   & \textbf{(0.812 - 0.837)} & \textbf{(0.906 - 0.909)} & \textbf{(0.86 - 0.868)}\\\hline
\end{tabular}
\caption{Comparison of $C_{index}$ performance tested on METABRIC and SEER (mean and 95\% confidence interval)}
\label{table:results}
\end{table}
Table \ref{table:results} displays the $C_{index}$ results of the experiments realized on SEER and METABRIC datasets. 
We can observe that, for METABRIC, DeepWeiSurv's performances exceed by far that of DeepHit and CPH. For the SEER data, DeepWeiSurv with $p$=1 outperfoms CPH (in BC and HD cases) and has a slight improvement over DeepHit especially for SEER HD data but without a significant difference (their confidence intervals did overlap). However, the improvement of  DeepWeiSurv with $p$=2 over all the other methods is highly statistically significant. We suspect that the good performances of DeepWeiSurv comes from its ability to learn implicitly the relationship between the covariates and the parameters without making any restrictive assumption.

\subsubsection{Censoring threshold sensitivity} 
\begin{figure}
    \centering
    \includegraphics[width=0.88\textwidth
    ]{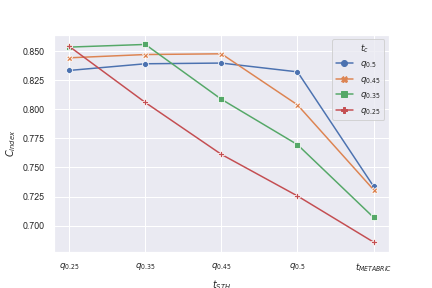}
    \centering
    \caption{The average of $C_{index}$ w.r.t survival time horizon $t_{STH}$ for every selected threshold $t_c$.}
    \label{scores_thresh}
\end{figure}
In the previous experiments the survival time horizon and the censoring threshold coincide, but it is not always the case. Since DeepWeiSurv predicts the conditional Weibull distributions with respect to the covariates, it is able to consider any survival time horizon given a censoring threshold. We add another experiment on METABRIC\footnote{We have chosen METABRIC dataset because of its small size compared to that of SEER dataset in order to avoid long calculations.} dataset where we assess DeepWeiSurv ($p=2$) performance with respect to censoring threshold time $t_c$. The aim of this experiment, is to check if DeepWeiSurv can handle data in highly censored setting for different survival time horizons. For this purpose, we apply the same experimental protocole as before, but changing the censoring threshold. We do this for some values of $t_c$ far below than that used in the previous experiment ($t_{c} = t_{METABRIC} = 8940$). This values, expressed in quantiles\footnote{We choose this values by using the quantiles of the survival times vector $T$.}, are carefully selected in order to have a significant added portion (compared to that of the adjacent value that precedes) of censored observations. As an observation may change from a censored status to an uncensored status by changing the threshold of censorship and vice versa, for each value of censoring threshold time $t_c$ we therefore have a new set of observed events $OE_{t_c} = \{(t_i, \delta_i) \|\delta_i = 1$ if $t_i < t_c$ else 0 \} (i.e. comparable events, and this contributes to the calculation of $C_{index}$). The training set, as it is selected, contains $ref = 866$ censored observations. Table \ref{thresh} gives the number of censored and uncensored observations of each selected value of $t_c$. For each value of $t_c$, we apply the 5-fold cross validation and then calculate the average $C_{index}$ for every survival time horizons $t_{STH}$. The results are displayed in Figure \ref{scores_thresh}.\\

\begin{table}
\centering
\captionsetup{justification=centering}
\begin{tabular}{ | c | c | c | c | c |}
\hline
{\textbf{$t_c$}}& 
& {No. uncensored} & {No. censored} & {Added portion (w.r.t $ref$)}\\ \hline \hline
 $q_{0.5}$&  &  1026  & 558 & 160\\
\hline
 $q_{0.45}$&  & 1127 &  457 & 261\\ \hline 
 $q_{0.35}$&  & 1248 &  336 & 382\\ \hline
 $q_{0.25}$&  & 1338 &  246 & 472\\
\hline
\end{tabular}
\caption{Distribution of training set's observations (censored/uncensored) for each selected censoring threshold.}
\label{thresh}
\end{table}
  Each curve in Figure \ref{scores_thresh} represents the scores calculated for a given censoring threshold $t_c$ in different survival time horizons $t_{STH}$ in x-axis. We can notice that the average score decreases when $t_c$ decreases which is expected because we have less and less of uncensored data which means that it becomes more and more difficult to model the distribution of survival times. However, DeepWeiSurv still performing well in highly censored setting. 

\section{Conclusion}
In this paper, we described a new approach, DeepWeiSurv, to the survival analysis. The key role of DeepWeiSurv is to predict the parameters of a mixture of Weibull distributions with respect to the covariates in presence of right-censored data. In addition to the fact that Weibull distributions are known to be a good representation for this kind of problem, it also permits to consider any survival time horizon given a censoring threshold.
Experiments on generated databases show that DeepWeiSurv converges to the real parameters when the survival time data follows a mixture of Weibull distributions whose parameters are a simple function of the covariates. On real datasets, DeepWeiSurv clearly outperforms the state-of-the-art approaches and demonstrates its ability to consider any survival time horizon. 

\bibliographystyle{plain}
\bibliography{refs}
\end{document}